\documentclass[letter,structabstract,longauth]{aa} 
\usepackage[dvips]{graphicx}
\usepackage{amsmath}
\usepackage{booktabs}
\usepackage{threeparttable}
\usepackage{wasysym}
\usepackage{natbib}
\usepackage{url}

\usepackage{txfonts}
\begin{document}
%
%   \title{The Dust, Ice and Gas in Time \emph{Herschel} Key Program First Results: Dust and Gas Spectroscopy of HD\,100546\thanks{\emph{Herschel} is an ESA space observatory with science instruments provided by European-led Principal Investigator consortia and with important participation from NASA.}
   \title{First results of the \emph{Herschel} Key Program `Dust, Ice and Gas in Time': Dust and Gas Spectroscopy of HD\,100546\thanks{\emph{Herschel} is an ESA space observatory with science instruments provided by European-led Principal Investigator consortia and with important participation from NASA.}
}

  \author{ B.~Sturm\inst{1} 
  \and J.~Bouwman\inst{1} 
  \and Th.~Henning\inst{1}
  \and N.~J.~Evans~II\inst{2} 
  \and B.~Acke\inst{10}\fnmsep\thanks{Postdoctoral Fellow of the Fund for  
  Scientific Research, Flanders}
  \and G.~D.~Mulders\inst{3,20}
  \and L.~B.~F.~M.~Waters\inst{3,10} 
  \and E.~F.~van~Dishoeck\inst{4,6}
  \and G.~Meeus \inst{5}
  \and J.~D.~Green \inst{2} 
  \and J.~C.~Augereau\inst{9}
  \and J.~Olofsson\inst{1} 
  \and C.~Salyk\inst{2}
  \and J.~Najita\inst{14}
\and G.~J.~Herczeg \inst{6} 
\and T.~A.~van~Kempen\inst{7} 
\and L.~E.~Kristensen \inst{4}
\and C.~Dominik\inst{3,21} 
\and J.~S.~Carr\inst{13} 
\and C.~Waelkens\inst{10} 
\and E.~Bergin\inst{17}
\and G.~A.~Blake\inst{12}
\and J.~M.~Brown \inst{6}
\and J.-H.~Chen\inst{2}
\and L.~Cieza\inst{16}\fnmsep\thanks{\emph{Spitzer} Fellow}
\and M.~M.~Dunham\inst{2}
\and A.~Glassgold\inst{18}
\and M.~G\"{u}del\inst{19}
\and P.~M.~Harvey\inst{2}
\and M.~R.~Hogerheijde\inst{4}
\and D.~Jaffe\inst{2}
\and J.~K.~J{\o}rgensen\inst{22}
\and H.~J.~Kim\inst{2}
\and C.~Knez\inst{15}
\and J.~H.~Lacy\inst{2}
\and J.-E.~Lee\inst{11}
\and S.~Maret\inst{9} 
\and R.~Meijerink \inst{4} 
\and B.~Mer\'{\i}n\inst{8} 
\and L.~Mundy\inst{15}
\and K.~M.~Pontoppidan\inst{12}
\and R.~Visser\inst{4}
\and U.~A.~Y{\i}ld{\i}z\inst{4}
          }

\institute{
  %1
  Max Planck Institute for Astronomy, K\"onigstuhl 17, D-69117 Heidelberg, Germany
  \and
  %2
  The University of Texas at Austin, Department of Astronomy, 1 University Station C1400, Austin, Texas 78712-0259, USA
  \and
  %3
  Astronomical Institute ``Anton Pannekoek'', University of Amsterdam, PO Box 94249, 1090 GE Amsterdam, The Netherlands
  \and
  %4
  Leiden Observatory, Leiden University, PO Box 9513, 2300 RA Leiden, The Netherlands
  \and
  %5
  Dpt. de F\'{i}sica Te\'{o}rica, Fac. de Ciencias, Universidad Aut\'{o}noma de Madrid, Campus Cantoblanco, 28049 Madrid, Spain
  \and
  %6
  Max Planck Institut f\"{u}r extraterrestrische Physik, Garching, Germany
  \and
  %7
  Harvard-Smithsonian Center for Astrophysics, 60 Garden Street, MS 42, Cambridge, MA 02138, USA
  \and
  %8
  Herschel Science Centre, European Space Astronomy Centre (ESA), P.O. Box 78, 28691 Villanueva de la Ca\~nada (Madrid), Spain
  \and 
  %9
  Laboratoire d'Astrophysique de Grenoble, CNRS/Universit\'{e} Joseph Fourier (UMR5571) BP 53, F-38041 Grenoble cedex 9, France
  \and
  %10
  Instituut voor Sterrenkunde, K.U.Leuven, Celestijnenlaan 200D, B-3001 Leuven, Belgium
  \and
  %11
  Department of Astronomy and Space Science, Astrophysical Research Center for the Structure and Evolution of the Cosmos, Sejong University, Seoul 143-747, Republic of Korea
  \and
  %12
  Caltech, Division of Geological \& Planetary Sciences, Mail Code 150-21, Pasadena, CA 91125
  \and
  %13
  Naval Research Laboratory, Code 7211, Washington, DC 20375, USA
  \and
  %14
  National Optical Astronomy Observatory, 950 N. Cherry Ave., Tucson, AZ 85719, USA
  \and
  %15
  Department of Astronomy, University of Maryland, College Park, MD 20742, USA
  \and
  %16
  Institute for Astronomy, University of Hawaii at Manoa, Honolulu, HI 96822, USA
  \and 
  %17
  Department of Astronomy, The University of Michigan, 500 Church Street, Ann Arbor, MI 48109-1042, USA
  \and 
  %18
  Astronomy Department, University of California, Berkeley, CA 94720, USA
  \and
  %19
  University of Vienna, Department of Astronomy, T\"urkenschanzstr. 17,  1180 Vienna, Austria
  \and
  %20
  SRON Netherlands Institute for Space Research, PO Box 800, 9700 AV,
  Groningen, The Netherlands
  \and
  %21
  Department of Astrophysics/IMAPP, Radboud University Nijmegen, P.O. Box 9010 6500 GL Nijmegen The Netherlands
  \and
  %22
  Centre for Star and Planet Formation, Natural History Museum of Denmark, 
  University of Copenhagen, {\O}ster Voldgade  5-7, DK-1350 Copenhagen K., Denmark
}

\offprints{B.~Sturm, \email{sturm@mpia.de}}

  \abstract
{We present far-infrared spectroscopic observations, taken with the Photodetector Array Camera and Spectrometer (PACS) on the \emph{Herschel Space Observatory}, of the protoplanetary disk around the pre--main--sequence star HD\,100546. These observations are the first within the DIGIT \emph{Herschel} key program, which aims to follow the evolution of dust, ice, and gas from young stellar objects still embedded in their parental molecular cloud core, through the final pre--main--sequence phases when the circumstellar disks are dissipated.}
{Our aim is to improve the constraints on temperature and chemical composition of the crystalline olivines in the disk of HD\,100546 and to give an inventory of the gas lines present in its far-infrared spectrum.
}
{The 69\,$\mu$m feature is analyzed in terms of position and shape to derive the dust temperature and composition. Furthermore, we detected 32 emission lines from five gaseous species and measured their line fluxes.
}
{The 69\,$\mu$m emission comes either from dust grains with $\sim$70~K at radii larger than 50\,AU, as suggested by blackbody fitting, or it arises from $\sim$200\,K dust at $\sim$13\,AU, close to the midplane, as supported by radiative transfer models. We also conclude that the forsterite crystals have few defects and contain at most a few percent iron by mass. Forbidden line emission from [C~{\sc ii}] at 157\,$\mu$m and [O~{\sc i}] at 63 and 145\,$\mu$m, most likely due to photodissociation by stellar photons, is detected.  Furthermore, five H$_2$O and several OH lines are detected. We also found high-J rotational transition lines of CO, with rotational temperatures of $\sim$300\,K for the transitions up to $J=22-21$ and $T\,\sim$800\,K for higher transitions.  }
{}

 \titlerunning{DIGIT First Results: Dust and Gas Spectroscopy of HD\,100546}
  \authorrunning{B. Sturm et al}
 
   \maketitle
%
%________________________________________________________________
%
\section{Introduction}
%

% something about disks
Circumstellar disks around young stars are the birthplaces of planetary systems. To understand planet formation, it is vital to study the processes that govern the evolution of gas and dust in these disks. PACS provides unique information in this field through observations of far-infrared (IR) solid-state features which are particularly sensitive to temperature and elemental composition. Moreover, PACS is well suited to study the warm gas of a few 100\,K, complementary to the hot gas probed in near-IR and the cold gas observed at millimeter wavelengths.

% specifics of HD 100546
An intensively studied pre-main-sequence star is the Herbig B9.5Vne star HD\,100546. While signs of ongoing accretion indicate the star's infancy \citep{Deleuil2004}, the estimated age of 10\,Myr makes it unusually old for a star with a disk \citep{vdAncker1997}. HD\,100546 is nearby (103 pc) and optical/near-IR scattered light imaging has revealed a wealth of structures in the disk \citep[e.g.,][]{Grady2001, Augereau2001, Ardila2007}. Based on the  mid-- to near-IR excess ratio, \citet{Bouwman2003} suggested an inner cavity in the disk. Later observations confirmed this gap \citep[e.g.,][]{Grady2005, Benisty2010}. This remarkable feature and the longevity of the HD\,100546 disk may point to a young planet in the inner 10 AU, making it a prime target for detailed studies.
%has revealed spiral arm structures in the outer disk at a few 100 AU

%dust
The system of HD\,100546 has a rich 2.4 -- 180~$\mu$m spectrum as observed with ISO \citep{Malfait1998}, showing a striking similarity with that of the comet Hale-Bopp \citep{Crovisier1997}. \citet{Juhasz2010} investigated the \emph{Spitzer} spectrum of the source. Strong forsterite emission is observed in both ISO and \emph{Spitzer} data. A continuum fit indicated crystalline dust at 210\,K and 50\,K \citep{Malfait1998}. 

%gas
The ISO spectrum shows a strong line of [O~{\sc i}] 63\,$\mu$m and a weaker [C~{\sc ii}] 158\,$\mu$m line. In the near-IR, \citet{Brittain2009} and \citet{vdPlas2009} observed CO ro--vibrational lines with a rotational temperature of $\sim$1000\,K and found no CO gas in the inner dust cavity. Observations of optical [O {\sc i}] and Balmer line emission, on the other hand, demonstrate that the inner disk is not completely devoid of gas \citep{AckevdAncker2006}. At submillimetre wavelengths, \citet{Panic2010} have detected pure rotational lines of CO up to $J=7-6$ probing a warm (60--70\,K) layer in the outer disk ($\sim$\,100\,AU). Information on the intermediate temperatures is still lacking.

% this paper:
Here we present an analysis of narrow features (about 1\,$\mu$m) in the far-IR spectrum of HD\,100546 over the full PACS range that was obtained within the \emph{Herschel} key program `Dust, Ice and Gas in Time' (DIGIT).
% Dust, Ice and Gas In Time (DIGIT) \emph{Herschel} Key Program. 

\section{Observation and data reduction}
\label{sec:datareduction_observation}
We used the spectral energy distribution (SED) mode, covering the wavelength range from 55 to 210~$\mu$m \citep[see ][for  details of the instrument and calibration]{Poglitsch2010}. Four up/down scans of the grating for two nod positions were performed with a chopper throw of 1$\arcmin$. Our data were processed in HIPE 3.0, provided by the \emph{Herschel} science center, up to level 1. We did not apply the integration ramp linearity, cross-talk and response drift corrections as these were not yet scientifically validated.

After level 1 processing, we extracted the spectra for the two nod positions separately from the central spaxel (spatial element). The wavelength--dependent point spread function (PSF) correction was taken from the PACS calibration document PICC-KL-TN-04$^1$.
%\footnote{ \url{http://herschel.esac.esa.int/Docs/Calibration/PACS_SpectroscopyPerformanceCalibration_11Mar2010.pdf}}. }
We first uniformly rebinned the spectra, using an over-sampling factor of 1 relative to the spectral resolution of the instrument and an up-sampling factor of 2, before combining the spectra of the two nod positions. The current flux calibration and relative spectral response function (RSRF) were derived from ground-based measurements. The correction factors, determined from the first in--space calibrations were applied (for details see the PACS calibration document PICC-KL-TN-041\footnote{http://herschel.esac.esa.int/Docs/Calibration /PACS\_SpectroscopyPerformanceCalibration\_11Mar2010.pdf}). 
We estimate that the absolute flux calibration is accurate  within 30\%. The RSRF is valid for spectral features not much wider than about 1~$\mu$m. For broader features, the RSRF is not yet well calibrated. Therefore, we will focus on spectral features narrower than about 1~$\mu$m. The wavelength calibration is accurate down to the spectral resolution as can be seen from the gas--phase lines, most of which are found within 0.02\,$\mu$m of their laboratory wavelength.
%From the same document the correction factors for the flux calibration, based on the first in--space calibration were taken.
%
\section{Results and Discussion}
\label{sec:results}
The PACS data confirm the tentative ISO detection of the 69~$\mu$m forsterite band first reported by \cite{Malfait1998} and deliver a band profile of very high quality (see Fig. \ref{fig:forsterite69_a}). The feature is characterized by a signal-to-noise ratio of about 10. In addition, we find a wealth of narrow molecular gas lines in the spectrum of HD\,100546, coming from CO, H$_2$O, and OH, and [C~{\sc ii}] and [O~{\sc i}] fine-structure lines (see Table \ref{tab:gas-lines}).

In this section we present an analysis of the 69 $\mu$m forsterite peak in terms of dust temperature and composition. In addition, we provide a compilation of the detected gas lines in the far-IR spectrum of HD\,100546 and derive the CO--rotational temperature. A full spectral decomposition, including a search for shallow and wide features (e.g. ice and hydrated silicate features),  will be presented in a follow-up paper.

\subsection{Forsterite}
\label{subsect:forsterite}
The `69~$\mu$m' band is a characteristic signature of forsterite - the Mg-rich end member of olivines \citep{Henning2010}. 
In order to determine the position and width of this feature, we fitted a Lorentz profile and a linear function for the continuum \citep[compare e.g.,][]{Koike2006} in the wavelength interval 68--71\,$\mu$m. The emission peaks at $69.198 \pm 0.008\, \mu$m and has a  FWHM of $0.64 \pm 0.03\,\mu$m. 
% if we show the Lorentz--fit, include (see Fig. \ref{fig:forsterite69_a}) before 'The emission peaks...'
% , with formula given by $\text{Mg}_{2-2x}\text{Fe}_{2x}\text{SiO}_4$, removed before \cite{Henning2010}

\begin{figure}
\centering
\includegraphics[scale=.85, angle=90]{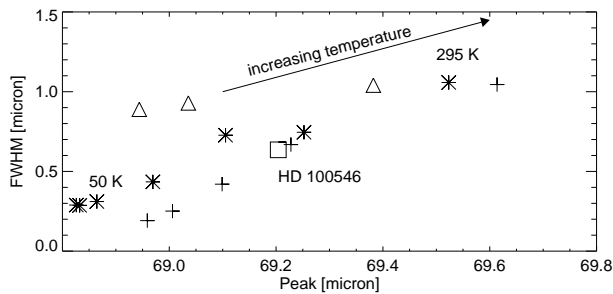}
\caption{\small{FWHM of the forsterite `69\,$\mu$m band' plotted over the peak position, fitted with a Lorentz profile. The square is our HD\,100546 data (size indicates error). For comparison, data from \citet[asterisks: `Kyoto sample', triangles: `Jena sample']{Koike2006}, \citet[crosses, DHS with 0.1\,$\mu$m grainsize]{Suto2006} are shown.}}
\label{fig:forsterite69_b}
\end{figure}

\begin{figure}
\centering
\includegraphics[scale=.85, angle=90]{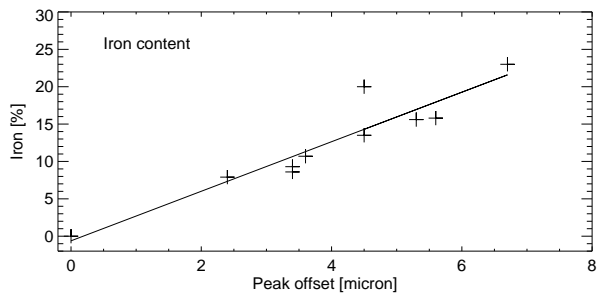}
\caption{\small{Effect of iron--rich olivines on the peak wavelength of the `69\,$\mu$m' band at room temperature. The offset in wavelength is relative to pure forsterite. The data are from \citet{Koike2003} and the  line is the best--fit linear function.}}
\label{fig:forsterite69_c}
\end{figure}

%Position and shape of the 69\,$\mu$m band depend on
The 69\,$\mu$m band profile depends on both the temperature and the iron content \citep[e.g.,][]{Koike2006, Koike2003} of the olivine crystals. Increasing the dust temperature leads to a shift of the peak position towards longer wavelengths and to an increase of the FWHM (Fig. \ref{fig:forsterite69_b}). Increasing the iron content in the olivine crystals has a very pronounced effect in shifting the band to longer wavelengths (Fig. \ref{fig:forsterite69_c}). Finally, the FWHM will increase with the number of lattice distortions \citep{Koike2010}.

Based on the peak position of the band, we can immediately conclude that the olivines cannot contain more than 3--4\% iron as this would cause the peak to appear at significantly longer wavelengths than observed. The PACS observations provide much better constraints on the iron content than the shorter-wavelength \emph{Spitzer} data \citep{Juhasz2010} because silicate bands in that spectral range are less sensitive to a change in Mg/Fe ratio. Furthermore, the small FWHM of the band profile indicates crystals with few defects \citep{Koike2010}. The small FWHM also shows that the dust emission comes from a narrow range of temperatures.

%
%---------------- one-col-figure of forsterite 69 micron ----------------------
\begin{figure}
\centering
\includegraphics[scale=.85]{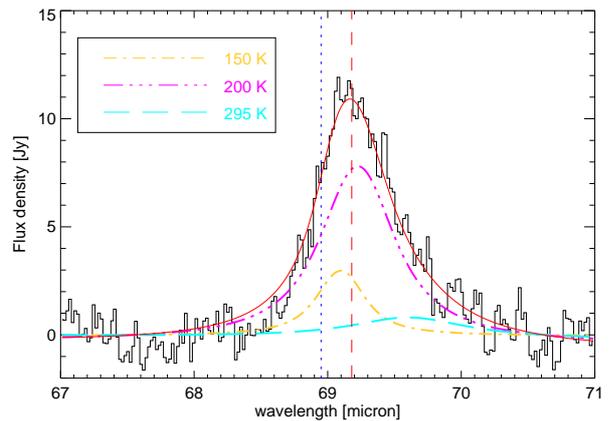}
\caption{\small{The continuum--subtracted forsterite emission at 69\,$\mu$m (histogram) overplotted with the best--fit model (red line) which is given by eq. (\ref{eq:dustmodel}). The contributions of the different temperatures are indicated. The dotted vertical line locates the peak position of 70\,K pure forsterite and the dashed one the position in the PACS spectrum, fitted with a Lorentz profile, which is not shown here.}}
\label{fig:forsterite69_a}
\end{figure}
%The dash--dotted lines represent the contributions of the laboratory samples at different temperatures.
%

Determining the temperature of the dust that produces the 69 $\mu$m feature from intrinsic properties of this band is more difficult because even a small amount of iron could balance a lower temperature. Assuming pure forsterite results in a dust temperature of about 200\,K  (Fig. \ref{fig:forsterite69_b}). To constrain the amount of colder dust, we fitted the observed profile with a weighted sum of laboratory data \citep{Suto2006} at different temperatures, ranging from 50--300~K (Fig. \ref{fig:forsterite69_a}). The measured flux $F_{\rm {68-71}\mu m}$ is given by
\begin{equation}
F_{\text{68-71}\mu\text{m}}(\lambda) = F_{\textnormal{cont}}(\lambda) + \sum_{i=1}^{5}{w_i \cdot \kappa(T_{\textnormal{dust}}^i, \lambda) \cdot B_{\lambda}(\lambda, T_{\textnormal{dust}}^i)}.
\label{eq:dustmodel}
\end{equation}
Here $F_{\rm{cont}}$ is the local continuum, and the $\kappa$ values are the mass absorption coefficients, computed from the optical constants of the \citet{Suto2006} forsterite sample, at the different dust temperatures $T_{\rm{dust}}^i$. Finally, the $w_i$ values are the relative weights of the flux at the different temperatures. In addition, we had to specify a dust shape distribution in order to calculate the mass absorption values. We used the DHS model \citep[distribution of hollow spheres,][]{Min2003} with a grain size of $0.1\,\mu$m, which has been found to be a good representation of observed silicate profiles \citep[e.g.,][]{Juhasz2010}. 
 
Under the assumption of pure forsterite, no more than a few percent of the dust emission can be coming from the 50 K component. This could be up to 9\%, if grains of $\sim$5\,$\mu$m radius are assumed. Such large grains can only exist in the deeper layers of the disk, where they are invisible to \emph{Spitzer}, as \citet{Juhasz2010} rule out grains $\ge$\,1\,$\mu$m. The bulk of the emission, under the assumption of pure forsterite, comes from $\sim$150--200\,K.

However, a temperature of 150 to 200\,K for the 69\,$\mu$m band would lead to tremendously higher fluxes in the 20--40\,$\mu$m region than observed by ISO and \emph{Spitzer}. Additionally, the predicted relative strength between the 24-- and 33\,$\mu$m forsterite bands would be inverted compared to the observations. In fact, temperatures of only 70\,K \citep{Bouwman2003} or even 50\,K \citep{Malfait1998} were derived from the ISO data.

These results can be reconciled in two ways. First, the dust producing the 69\,$\mu$m feature is at 70--50\,K. The shift from the band position in the 70--50\,K laboratory data to the one observed by PACS could be caused by an admixture of iron. Given the information in Fig. \ref{fig:forsterite69_c}, we need at least 2 but no more than 3\% for such a shift, considering that the data were taken at $\sim$300\,K.

An alternative explanation is based on the optical depth of the disk which decreases with increasing wavelength. The $\tau = 1$ surface is deep into the disk at 69\,$\mu$m, even reaching the midplane \citep{Mulders2010}. The disk of HD\,100546 is a special case in this respect: because of its large inner hole, the midplane temperature at 13\,AU is much higher ($\sim$150--200\,K) than in disks without a hole so that the far-infrared features can be produced in heated layers close to the midplane from which the radiation at 70\,$\mu$m can escape. In the most extreme case the emission could arise from pure forsterite of 200\,K temperature, located in the midplane at 13\,AU.

In this scenario the location of the forsterite near the inner edge of the disk supports the crystallization by collisional cascades or shocks induced by the suggested giant planet.

%
%
%---------------- forsterite 69 micron FWHM over peak ----------------------
%

%---------------- forsterite 69 micron Fe_(x) over peak -----------------------

\subsection{Gas lines}
\label{subsect:gaslines}
\begin{figure}
\centering
\includegraphics[width=0.5 \textwidth]{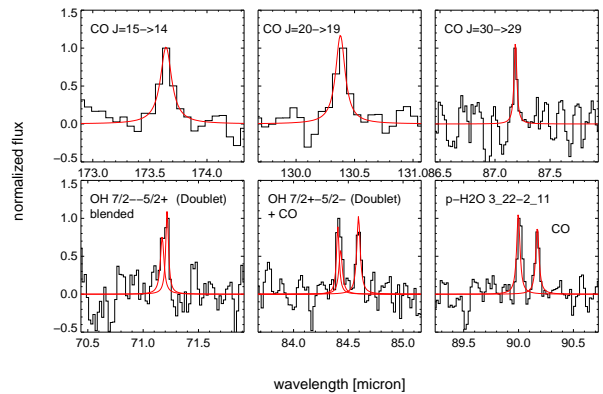}
\caption{\small{Selected gas lines from the PACS spectrum of HD\,100546. The continuum--subtracted, normalized fluxes are plotted as histograms. The continuum has been fitted with a linear function of the wavelength and the peaks with Lorentz profiles. The second peak in the H$_2$O 90\,$\mu$m panel is CO $J=$29--28.}}
\label{fig:gas-lines}
\end{figure}

We identified 32 gas lines (Table \ref{tab:gas-lines}) from the ISO line lists\footnote{\url{http://www.mpe.mpg.de/ISO/linelists/index.html}}. CO lines up to the $J$=31--30 transition were detected as well as H$_2$O and OH lines. Selected molecular lines are shown in Fig.~\ref{fig:gas-lines}. Furthermore, we confirm the ISO detections of [O\,{\sc i}] and [C\,{\sc ii}] fine-structure lines \citep{Lorenzetti2002}. The integrated line fluxes, obtained from fitted Lorentz profiles, are in Table \ref{tab:gas-lines}. The $\text{[O~{\sc i}]}$ line fluxes agree within 30\% with those seen by ISO \citep{Lorenzetti2002}, while the $\text{[C~{\sc ii}]}$ line flux is about a factor of 3 weaker. Significant flux outside the pointsource PSF is detected for the [C\,{\sc ii}] line, which is not seen for any other line, suggesting that extended emission dominates the larger ISO beam. Through comparison of the flux ratios $\text{[O {\sc i}]}_{63} / \text{[O {\sc i}]}_{145}$ and $\text{[O {\sc i}]}_{63} / \text{[C {\sc ii}]}_{158}$ with the models from \citet{Kamp2010} a gas mass of $\sim10^{-3}$\,M$_{\footnotesize{\astrosun}}$ is derived.

According to our knowledge, the detection of H$_2$O is one of the first in the disk of a Herbig Ae/Be star. The emission seems to emerge from optically thick gas. A possible source for the H$_2$O could be sublimation of water ice. A detailed analysis will be presented in a future paper.

%
%------------------- two-col figure of some gas line profiles -----------------
%
%
%-------------------- one-col table of gas line fluxes etc -------------------
\begin{table}%[h]
\begin{center}
\begin{threeparttable}
\begin{tabular}{@{}rrrr@{}}
\toprule
Species    &   Transition  & Peak                    & Lineflux \\
        &               &  [$\mu\textnormal{m}$]  & $10^{-17}[\text{W/m}^2]$\\
\cmidrule{1-4}
$\text{[O~{\sc i}]}$ &      $^3\text{P}_1-^3\text{P}_2$ &   63.19 & $ 554.37\,\pm\, 5.0$\\
$\text{[O~{\sc i}]}$ &      $^3\text{P}_0-^3\text{P}_1$ &  145.54 & $  35.70\,\pm\, 1.3$\\
$\text{[C~{\sc ii}]}$ &  $^2\text{P}_{3/2}-^2\text{P}_{1/2}$ &  157.75 & $ 31.87\,\pm\, 1.0$\\
CO &     $J=14-13 $&  186.01 & $   7.42\,\pm\, 0.9$\\
CO &     $J=15-14 $&  173.64 & $  11.50\,\pm\, 0.8$\\
CO &     $J=16-15 $&  162.82 & $   8.25\,\pm\, 0.9$\\
CO &     $J=17-16 $&  153.28 & $  10.49\,\pm\, 0.8$\\
CO &     $J=18-17 $&  144.79 & $   9.85\,\pm\, 1.0$\\
CO &     $J=19-18 $&  137.21 & $   8.86\,\pm\, 0.7$\\
CO &     $J=20-19 $&  130.38 & $   7.10\,\pm\, 0.7$\\
CO &     $J=21-20 $&  124.18 & $   5.94\,\pm\, 0.6$\\
CO &     $J=22-21 $&  118.59 & $   5.92\,\pm\, 0.9$\\
CO\tnote{1}&     $J=23-22 $&  113.47 & $   4.76\,\pm\, 1.2$\\
CO &     $J=24-23 $&  108.77 & $  10.56\,\pm\, 1.3$\\
CO &     $J=26-25 $&  100.47 & $  12.71\,\pm\,  2.4$\\
CO &     $J=27-26 $&   96.78 & $  10.20\,\pm\, 1.4$\\
CO &     $J=28-27 $&   93.35 & $  10.00\,\pm\,  1.0$\\
CO &     $J=29-28 $&   90.17 & $  11.83\,\pm\, 1.6$\\
CO &     $J=30-29 $&   87.19 & $   8.46\,\pm\, 1.0$\\
CO\tnote{1} &     $J=31-30 $&   84.40 & $   4.90\,\pm\, 1.5$\\ %OH dublet model
OH    & 3/2,~ $9/2^+-7/2^-$ &   65.29 & $  10.00\,\pm\, 1.5$\\
OH\tnote{1} &1/2,~ $7/2^--5/2^+$ &   71.17 & $   9.17\,\pm\, 1.4$\\
OH\tnote{1} &1/2,~ $7/2^+-5/2^-$ &   71.22 & $  10.10\,\pm\, 1.1$\\
OH\tnote{1} &3/2,~ $7/2^+-5/2^-$ &   84.42 & $   14.96\,\pm\, 1.7$\\
OH    &3/2,~ $7/2^--5/2^+$ &   84.59 & $  14.96\,\pm\, 1.5$\\
OH    &1/2,~ $5/2^+-3/2^-$ &   98.73 & $   8.74\,\pm\, 1.5$\\
OH    &1/2,~ $3/2^--1/2^+$ &  163.41 & $   2.60\,\pm\, 0.6$\\
p-H2O &   $ 3_{22}-2_{11} $ &   90.00 & $  14.32\,\pm\, 1.2$\\
o-H2O\tnote{1} &   $ 4_{14}-3_{03} $ &  113.53 & $   3.64\,\pm\, 0.9$\\
o-H2O &   $ 3_{30}-3_{21} $ &  136.48 & $   4.52\,\pm\, 0.7$\\
p-H2O &   $ 3_{13}-2_{02} $ &  138.54 & $   3.17\,\pm\, 0.7$\\
o-H2O &   $ 2_{12}-1_{01} $ &  179.54 & $   5.53\,\pm\, 0.8$\\

\bottomrule
\end{tabular}
\begin{tablenotes}
  \item[1] \footnotesize{deblended lines, CO not used in rotational diagram}
\end{tablenotes}
\end{threeparttable}
\caption{\small{Peak wavelengths and line fluxes for the identified gas lines between 60 and 190\,$\mu$m. Only the formal errors are given, not including systematic errors due to the PACS spectral response function.}}
\label{tab:gas-lines}
\end{center}
\end{table}

\begin{figure}
\centering
\includegraphics[scale=.85]{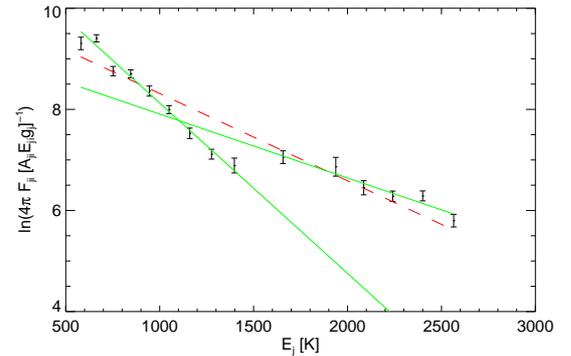}
\caption{\small{Rotational diagram of the CO lines in HD\,100546 (crosses), overplotted with our models, using eq. 2 from \citet{Justtanont2000}. A single temperature fit (dashed) gives reasonable results, but a two--temperature approach (solid) yields a better description of the data. CO $J$=23--22 and $J$=31--30 are blended with H$_2$O and OH respectively and were not used in the fit, nor are they shown here.}}
\label{fig:rotational_diagram}
\end{figure}

Our focus here is on the newly detected CO lines. In Fig. \ref{fig:rotational_diagram} we show the rotational diagram of the CO lines. Although a single-temperature $T$\,=\,$580 \pm 14$\,K provides a reasonable fit (reduced $\chi^2 = 15$) to the data, a better representation could be obtained by a two--temperature model (reduced $\chi^2 = 1.8$). The transitions up to $J$=22--21 are fitted by $T$\,=\,$300 \pm 12$\,K. The higher rotational transitions require a hotter gas component of $800 \pm 100$\,K providing the connection to the high-temperature gas in the innermost part of the disk detected by near-infrared spectroscopy \citep{Brittain2009, vdPlas2009}. This inferred temperature range is consistent with CO in a heated surface layer of the disk as found in disk models that do not assume equal gas and dust temperatures but include an explicit treatment of the heating and cooling processes \citep[e.g.,][]{Kamp2004}. The 300\,K component could also emerge from the region in the midplane close to the gap.
%Jonkheid2007 after Kamp2004
%and the inferred temperature range is consistent with that found for CO in disk models that do not take the gas temperature to be equal to that of the dust but include an explicit treatment of the heating and cooling processes \citep[e.g.,][]{Kamp2004, Jonkheid2007}.
    
According to \citet{Brittain2009} the $4.7\,\mu$m CO fundamental emission shows both a UV fluorescent and a thermal component. The latter was fitted with a column density of at least $2.3 \cdot 10^{15}$\,cm$^{-2}$ \citep[for discussion, see][]{Brittain2009}. The submillimeter data give a typical CO column density of $10^{18}$ cm$^{-2}$ \citep{Panic2010}. From Fig. \ref{fig:rotational_diagram}, assuming that the emission originates from the area between 13 and 100\,AU, we derive column densities of $10^{16}$ to $10^{17}$\,cm$^{-2}$ in both models. These values indicate that PACS probes the part of the disk between the near-IR and submillimeter data. 

%
%---------------- one-col-figure of rotational diagram ----------------------

\section{Conclusions}
\label{sec:discussion}
The PACS data provide the strongest constraints yet on the composition of olivines in a protoplanetary disk. We found the olivines to be extremely iron-poor (less than 3--4\% iron). Our observations can be modeled with dust at $\sim$70\,K, consistent with ISO results, but requiring at least 2\% iron. On the other hand the determination of the temperature of the forsterite based on intrinsic features of the 69\,$\mu$m band instead of using a continuum analysis offers a second option: The emission may emerge from pure forsterite at 200\,K and 13\,AU close to the midplane. This region is optically thick for shorter wavelengths.

The far-infrared spectrum of HD\,100546 contains a wealth of molecular gas lines including CO, H$_2$O, and OH. We found CO transitions up to $J$=31--30 which arise from gas in the temperature range of 300 to 800 K and appear to sample gas in the surface layers of the disk at temperatures and surface densities intermediate to those probed by submillimeter and near-infrared CO measurements.

%Acknowledgement
\begin{acknowledgements}
Support for this work, part of the \emph{Herschel} Open Time key program, was provided by NASA through an award issued by the Jet Propulsion Laboratory, California Institute of Technology. G. Meeus is partly supported by Spanish grant AYA 2008-01727. The authors thank the referee C.~Grady for helpful comments that improved the paper.
\end{acknowledgements}

\bibliographystyle{aa}
\bibliography{hd100546}

\end{document}